\documentstyle[epsfig,longtable]{aipproc}

\begin{document}
\title{Neutrino Physics at Muon Colliders}

\author{Bruce J. King}
\address{Brookhaven National Laboratory\\
email: bking@bnl.gov}\thanks{Presented at the Fourth International Conference
on the Physics Potential and Development of Muon Colliders, San Francisco,
December 10-12, 1997. This work was performed under the auspices of
the U.S. Department of Energy under contract no. DE-AC02-76CH00016.}
\maketitle

\begin{abstract}
  An overview is given of the neutrino physics potential of future muon
storage rings that use muon collider technology to produce, accelerate and
store large currents of muons.
\end{abstract}

\section*{Introduction}
%%%%%%%%%%%%%%%%%%%%%%%

  This paper gives an overview of the neutrino physics
possibilities at a future muon storage ring, which can be either a muon
collider ring or a ring dedicated to neutrino physics that uses muon
collider technology to store large muon currents.
It summarizes a previous more detailed description of these topics
by this author~\cite{nufnal97}.

   After a general characterization of the neutrino beam and its interactions,
some crude quantitative estimates are given for the physics performance of
a muon ring neutrino experiment (MURINE) consisting of a high rate, high
performance neutrino detector at a 250 GeV muon collider storage ring.

\section*{Neutrino Production and Event Rates}
%%%%%%%%%%%%%%%%%%%%%%%%%%%%%%%%%%%%%%%%%%%%%%

  Neutrinos are emitted from the decay of muons in the collider ring:
\begin{eqnarray}
\mu^- & \rightarrow & \nu_\mu + \overline{\nu_{\rm e}} + {\rm e}^-,
                                             \nonumber \\
\mu^+ & \rightarrow & \overline{\nu_\mu} + \nu_{\rm e} + {\rm e}^+.
                                                 \label{eq:nuprod}
\end{eqnarray}

   The thin pencil beams of neutrinos for experiments will be produced
from long straight sections in either the collider ring or a
ring dedicated to neutrino physics. From relativistic kinematics, the
forward hemisphere in the muon rest frame will be boosted, in the lab
frame, into a narrow cone with a characteristic opening half-angle,
$\theta_\nu$,
given in obvious notation by
\begin{equation}
\theta_\nu \simeq \sin \theta_\nu = 1/\gamma =
\frac{m_\mu}{E_\mu} \simeq \frac{10^{-4}}{E_\mu ({\rm TeV})}.
                                                   \label{eq:thetanu}
\end{equation}

  The large muon currents and tight collimation of the neutrinos results
in extremely intense beams -- intense enough even to constitute a potential
off-site radiation hazard~\cite{bjkrad}.

For the example of 250 GeV muons, the neutrino beam will have an opening
half-angle of approximately 0.4 mrad. The final focus regions around collider
experiments are important exceptions to equation~\ref{eq:thetanu} since
the muon beam itself will have an angular divergence in these regions that
is large enough to significantly spread out the neutrino beam.

%..# interactions in a detector (2 equations)

  For TeV-scale neutrinos, the neutrino cross-section is approximately
proportional to the neutrino
energy, $E_\nu$. The charged current (CC) and
neutral current (NC) interaction cross sections for neutrinos and
antineutrinos have numerical values of~\cite{quigg}:
\begin{equation}
 {\rm \sigma_{\nu N}\; for\;}
 \left(
 \begin{array}{c}
   \nu_-CC \\
   \nu_-NC \\
   \overline{\nu}-CC \\
   \overline{\nu}-NC
 \end{array}
 \right)\;
 \simeq
 \left(
 \begin{array}{c}
    0.72 \\ 0.23 \\ 0.38 \\ 0.13
   \end{array}
 \right)
\times {\rm \frac{E_\nu}{1\:TeV}}
\times 10^{-35}\: {\rm cm^2}.
                                            \label{eq:xsec}
\end{equation}

  These cross sections are easily converted into approximate experimental
event rates for the example used in reference~\cite{nufnal97} of a
250+250 GeV collider with a 200 meter straight section.
For a general purpose detector subtending the boosted forward hemisphere
of the neutrino beam:
\begin{equation}
 {\rm Number\; of\;}
 \left(
 \begin{array}{c}
   \nu_\mu-CC \\
   \nu_\mu-NC \\
   \overline{\nu}_e-CC \\
   \overline{\nu}_e-NC
 \end{array}
 \right)\;
 {\rm events/yr} \simeq
 \left(
 \begin{array}{c}
    2.6 \\ 0.8 \\ 1.4 \\ 0.5
   \end{array}
 \right)
\times 10^7 \times l [{\rm g.cm^{-2}}],
                                            \label{eq:genevents}
\end{equation}
where $l$ is the detector length.
For a long baseline detector in the center of the neutrino beam:
\begin{equation}
 {\rm Number\; of\;}
 \left(
 \begin{array}{c}
   \nu_\mu-CC \\
   \nu_\mu-NC \\
   \overline{\nu}_e-CC \\
   \overline{\nu}_e-NC
 \end{array}
 \right)\;
 {\rm events/yr} \simeq
 \left(
 \begin{array}{c}
    1.4 \\ 0.4 \\ 0.7 \\ 0.2
   \end{array}
 \right)
\times 10^7 \times {\rm \frac{M[kg]}{(L[km])^2} },
                                            \label{eq:longevents}
\end{equation}
where $M$ is the detector mass and $L$ the distance from the neutrino
source.

These event rates are several orders of magnitude higher than in
today's neutrino beams from accelerators.

\section*{A General Purpose Neutrino Detector}
%%%%%%%%%%%%%%%%%%%%%%%%%%%%%%%%%%%%%%%%%%%%%%

%..detector figure
%..table of subdetectors
%..overview of subdetectors

%...*** neutrino detector figure ***
\begin{figure}[t!] %
\centering
\includegraphics[height=3.5in,width=3.5in]{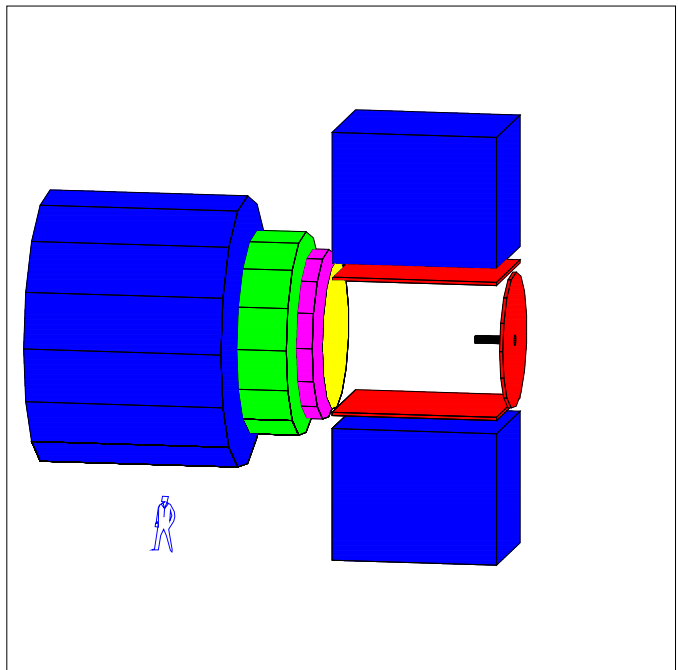}
\caption{Example of a general purpose neutrino detector. A human figure in
the lower left corner illustrates its size. The neutrino target is the small
horizontal cylinder at mid-height on the right hand side of the detector. Its
radial extent corresponds roughly to the radial spread of the neutrino pencil
beam, which is incident from the right hand side. Further details are given
in the text.}
\label{detector_fig}
\end{figure}

Figure~\ref{detector_fig} is an example of the sort of high rate general
purpose neutrino detector that would be well matched to the intense neutrino
beams. Note the contrast with the kilotonne-scale calorimetric targets used
in today's high rate neutrino experiments.

The neutrino target is a 1 meter long stack of CCD tracking planes with
a radius of 10 cm chosen to
match the beam radius at approximately 200 meters from
production for a 250 GeV muon beam.
 It contains 750 planes of 300 micron thick silicon CCD's, corresponding to
a mass per unit area of approximately 50 ${\rm g.cm^{-2}}$, about
2.5 radiation lengths and 0.5 interaction lengths.
Equation~\ref{eq:genevents} predicts a very healthy $2 \times 10^9$ CC
interactions per year for this target.

  Besides providing the mass for neutrino
interactions, the tracking target allows precise reconstruction of the event
topologies from charged tracks, including event-by-event vertex tagging
of those events containing charm or beauty hadrons or tau
leptons. Given the favorable vertexing geometry and the few-micron typical
CCD hit resolutions, it is reasonable to expect almost 100 percent
efficiency for b tagging, perhaps 70 to 90 percent efficiency for
charm tagging and excellent discrimination between b and c decays.

 The target in figure~\ref{detector_fig} is surrounded by a
time projection chamber (TPC) tracker in a vertical
dipole magnetic field. The
characteristic dE/dx signatures from the tracks would identify
each charged particle. Further particle ID is provided by the Cherenkov
photons that are produced in the TPC gas then reflected by a spherical
mirror at the downstream end of the tracker and focused onto a read-out
plane at the upstream end of the target.
The mirror is backed by electromagnetic and hadronic calorimeters
and, lastly, by iron-core toroidal magnets for muon ID.

\section*{Neutrino Interactions and Their Experimental Interpretation}
%%%%%%%%%%%%%%%%%%%%%%%%%%%%%%%%%%%%%%%%%%%%%%%%%%%%%%%%%%%%%%%%%%%%%%

  The dominant interaction of TeV-scale neutrinos is deep inelastic
scattering (DIS) off nucleons (i.e. protons and neutrons).
 There are 2 types of DIS: neutral current (NC) and charged
current (CC) scattering.
In neutral current (NC) scattering, the neutrino is deflected by
a nucleon ($N$) and loses energy with the production of
several hadrons ($X$):
\begin{equation}
\nu + N \rightarrow \nu + X,
                                        \label{eq:ncN}
\end{equation}
This comprises about 25 percent of the total cross section
and is interpreted as elastic scattering off one of the many quarks inside the
nucleon through the exchange of a virtual neutral Z boson:
\begin{equation}
\nu + q \rightarrow \nu + q.
                                        \label{eq:ncq}
\end{equation}

  Charged current (CC) scattering is similar to NC scattering
except that the neutrino turns into its corresponding charged lepton:
\begin{eqnarray}
\nu + N & \rightarrow & l^- + X, \nonumber \\
\overline{\nu} + N & \rightarrow & l^+ + X,
                                        \label{eq:ccN}
\end{eqnarray}
where $l$ is an
electron/muon for electron/muon neutrinos. At the more fundamental quark
level a charged W boson is exchanged with a quark ($q$), which is turned into
another quark species ($q'$) whose charge differs by one unit.
\begin{eqnarray}
\nu + q & \rightarrow & l^- + q', \nonumber \\
\overline{\nu} + q' & \rightarrow & l^+ + q.
                                        \label{eq:ccq}
\end{eqnarray}

  The relativistically invariant quantities that are routinely extracted in
DIS experiments are 1) Feynman $x$, the fraction of the nucleon
momentum
carried by the struck quark, 2) the inelasticity, $y = E_{\rm hadronic}/E_\nu$,
which is related to the scattering angle of the neutrino in the neutrino-quark
CoM frame, and 3) the momentum-transfer-squared,
$Q^2 = 2 M_{proton} E_\nu x y$. MURINE's will have the further capability
of reconstructing the hadronic 4-vector, resulting in a much better
characterization of each interaction.

  The final state quark always ``hadronizes'' at the nuclear distance scale,
combining with quark-antiquark pairs to produce the several hadrons seen in
the detector. Final state c and b quarks can be identified by vertex tagging
of the decaying charm or beauty hadrons that contain them, and
some statistically based flavor tagging will also be available for u, d or s
final state quarks, using so-called ``leading particle
effect''~\cite{nufnal97}.

\section*{Physics Opportunities}
%%%%%%%%%%%%%%%%%%%%%%%%%%%%%%%%

 Neutrino interactions are interesting both in their own right and as probes
of the quark content of nucleons, so a MURINE has wide-ranging potential to
make advances in many areas of elementary particle physics. This section gives
an overview for measurements involving the CKM quark mixing matrix,
nucleon structure and QCD, electroweak measurements,
neutrino oscillations and, finally, studies
of charmed hadrons.

  There is considerable theoretical interest in the mixture of final state
quarks produced in CC interactions. The struck quark can be converted into
any of the three final state quarks that differ by one unit of charge:
a down (d), strange (s), or bottom(b) quark can be converted into an up (u),
charmed (c), or top (t) quark and vice versa. In practice, production of the
heavy top quark is kinematically forbidden at these energies and the production
of other quark flavors is influenced by their mass. Beyond this, the Standard
Model of elementary particle physics (SM) predicts the probability for the
interaction to be proportional to
the absolute square of the appropriate element in the so-called
Cabbibo-Kobayashi-Maskawa (CKM) quark mixing matrix, a unitary matrix
with 4 free parameters whose values are not predicted by the SM. 
Improved measurements involving the
CKM matrix will test the SM hypothesis. It is of particular interest that
one of the 4 parameters is a complex phase that is postulated as an
explanation for CP violation -- the intriguing experimental phenomenon that
particles may have tiny deviations from the properties that mirror those of
their antiparticles.

 The experimentally determined values for the 9 mixing probabilities
are given in table~\ref{ckm_table}~\cite{ckm}, along with their current
percentage uncertainties and speculative projections~\cite{nufnal97} 
for how the uncertainties could be reduced by a MURINE. From the
large improvements in 4 of the 9 uncertainties
it is clear that a MURINE has potential for tremendous
improvements in measuring the quark mixing matrix, and more detailed studies
are clearly desirable.

\begin{table}[ht!]
\caption{
Quark mixing probabilities. Threshold suppression
due to quark masses has been neglected. In practice, this will reduce
the mixing probabilities to the heavier c and b quarks to below the
values given in the table and will prevent any mixing to the top quark.
The second row for each quark gives
current percentage uncertainties in quark mixing probabilities
and speculative projections of the uncertainties after analyses from
a MURINE. The two uncertainties in brackets have not been measured directly
from tree level processes. The uncertainties assume that no unitarity
constraints have been used.}
\begin{tabular}{|c|lll|}
\hline
          & \hspace{0.2 cm} \bf{d} & \hspace{0.2 cm} \bf{s}  &
                                          \hspace{0.3 cm}\bf{b}  \\
\hline
\bf{u}    &   \bf{0.95}  &  \bf{0.05}    &  \bf{0.00001}  \\
          &   $\pm$0.1\%  &  $\pm$1.6\%    & $\pm$50\% $\rightarrow$ 1-2\% \\
          &&& \\
\bf{c}    &   \bf{0.05}  &  \bf{0.95}    &  \bf{0.002}  \\
          &   $\pm$15\% $\rightarrow$ 0.2-0.5\%   &
              $\pm$35\% $\rightarrow$ $\sim 1$\%         &
              $\pm$15\% $\rightarrow$ 3-5\%     \\
          &&& \\
\bf{t}    &  \bf{0.0001}  &  \bf{0.001}  &  \bf{1.0} \\
          &   ($\pm$25\%)   &  ($\pm$40\%)   & $\pm$30\%
\label{ckm_table}
\end{tabular}
\end{table}

  Another major motivation for MURINE's is the potential for greatly improved
measurements of nucleon structure functions -- the momentum distributions of
quarks inside the nucleon. This provides~\cite{deltamw}
important tests of
quantum chromodynamics (QCD) -- the theory of the strong interaction that is
widely accepted for its elegance and simplicity but which has not been
experimentally verified at the level of the electroweak theory. A MURINE
might well be the best single experiment of any sort for the examination of
perturbative QCD~\cite{nufnal97}.

  Neutrino physics has also had an important historical role in measuring
the electroweak mixing angle, which is simply related to the mass ratio of
the W and Z intermediate vector bosons: 
\begin{equation}
\sin^2\theta_W \equiv 1 - \left( \frac{M_W}{M_Z} \right) ^2.
                        \label{eq:wma}
\end{equation}
  Now that $M_Z$ has been precisely measured at LEP, measurements of
$\sin^2\theta_W$ in neutrino physics can be directly converted to
predictions for
the W mass. The comparison of this prediction with direct $M_W$
measurements in collider experiments constitutes a precise prediction
of the SM and a sensitive test for exotic physics modifications to
the SM~\cite{deltamw}.
Reference~\cite{nufnal97} estimates that the
predicted uncertainty in $M_W$ from a MURINE analysis might be of order
10 MeV, which improves by more than an order of magnitude on
today's neutrino experiments~\cite{deltamw,ccfrwma} and is
comparable with the projected best direct measurements from future
collider experiments.

  A neutrino property that is currently drawing much interest is the question
of whether neutrinos have a non-zero mass. If they do then it is possible
for the 3 neutrino flavors to mix, perhaps producing neutrino oscillations that
can be observed using a neutrino beam. The probability for an oscillation
between two of the flavors is given by\cite{pdg}:

\begin{equation}
{\rm Oscillation\; Probability} = \sin ^2 \theta \times
       \sin ^2 \left( 1.27 \frac{\Delta m^2 [{\rm eV^2}].L[km]}
                                            {E_\nu [GeV]} \right),
                                        \label{eq:oscprob}
\end{equation}

where the first term gives the mixing strength and the second term gives the
distance dependence.

Reference~\cite{nufnal97} obtains
the following order-of-magnitude mass limit for an assumed long-baseline
detector with reasonable parameters and with full mixing:
\begin{equation}
\Delta m^2 |_{min} \sim O(10^{-4})\: eV^2,
                                        \label{eq:deltamsq}
\end{equation}
independent of the distance to the detector.
Similarly, a mixing probability sensitivity
for $10^{10}$ events in a short-baseline detector is found to be as low as
\begin{equation}
\sin ^2 \theta |_{min} \sim O(10^{-7}),
                                        \label{eq:thetaosc}
\end{equation}
for the most favorable value of $\Delta m^2$.
Both of these estimates apply generically to all
3 possible mixings between 2 flavors:
$\nu_e \leftrightarrow \nu_\mu$,
$\nu_e \leftrightarrow \nu_\tau$ and
$\nu_\mu \leftrightarrow \nu_\tau$.
(See also reference~\cite{Geer} for another discussion of neutrino
oscillations at a MURINE.)

The $\Delta m^2$ estimate is more than an order
of magnitude better than any proposed accelerator or reactor experiments for
$\nu_\mu \leftrightarrow \nu_\tau$
and $\nu_e \leftrightarrow \nu_\tau$, and is competitive with the best such
proposed experiments for $\nu_e \leftrightarrow \nu_\mu$.
The estimated value for $\sin ^2 \theta |_{min}$ is even more impressive
-- orders of magnitude better than in any other current or
proposed experiment for each of the three possible oscillation.

  As an interesting final topic, MURINE's should be rather impressive
factories for the study of charm -- with
a clean, well reconstructed sample of several times $10^8$ charmed hadrons
produced in $10^{10}$ neutrino interactions.
There are several interesting physics motivations for charm
studies at a MURINE~\cite{charmphysics}. As an example, particle-antiparticle
mixing has yet to be observed in the charm sector~\cite{d0mixing}, and 
it is quite plausible~\cite{nufnal97}
that a MURINE would provide the first observation of
${\rm D^0 - \overline{D^0}}$ mixing.

\section*{Summary}
%%%%%%%%%%%%%%%%%%

   The intense neutrino beams at muon collider complexes should
usher in an exciting new era of neutrino physics experiments, with
great advances expected in both traditional and new areas of neutrino
physics.


\begin{references}
\bibitem{nufnal97} B.J. King,
    {\it Neutrino Physics at a Muon Collider},
    Proc. Workshop on Physics at the First Muon Collider
and Front End of a Muon Collider, Fermilab, November 6-9, 1997.
\bibitem{bjkrad} B.J. King,
    {\it Assessment of the prospects for muon colliders},
    paper submitted in partial fulfillment of requirements
    for Ph.D., Columbia University, New York (1994);
    B.J. King,
    {\it A Characterization of the Neutrino-Induced
    Radiation Hazard at TeV-Scale Muon Colliders},
    BNL Center for Accelerator Physics internal report 162-MUON-97R,
    to be submitted for publication.
\bibitem{quigg} See, for example, Chris Quigg,
    {\it Neutrino Interaction Cross Sections}, FERMILAB-Conf-97/158-T.
\bibitem{ckm} Values extracted from Andrzej J. Buras,
    {\it CKM Matrix: Present and Future}, TUM-HEP-299/97.
\bibitem{deltamw}  Janet M. Conrad, Michael H. Shaevitz and Tim Bolton,
    {\it Precision Measurements with High Energy Neutrino Beams},
    hep-ex/9707015, submitted to Rev. Mod. Phys. (1997)
\bibitem{ccfrwma}  K.S. McFarland {\it et al.} (CCFR/NuTeV Collaboration)
    {\it A Precision Measurement of Electroweak Parameters in
    Neutrino-Nucleon Scattering}, FNAL-Pub-97/001-E.
    B.J. King, Columbia University Ph.D. Thesis, 1994;
    Nevis Report: Nevis-283, CU-390, Nevis Preprint R-1500 (1994).
\bibitem{pdg}
    R.M. Barnett et al., Physical Review D54, 1 (1996) 
    and 1997 off-year partial update for the 1998 edition available on 
    the PDG WWW pages (URL: http://pdg.lbl.gov/).
\bibitem{Geer}   S. Geer,
    {\it The Physics Potential of Neutrino Beams From Muon Storage Rings},
    Proc. Workshop on Physics at the First Muon Collider
    and Front End of a Muon Collider, Fermilab, November 6-9, 1997.
\bibitem{charmphysics}
    I.I Bigi, {\it Open Questions in Charm Decays Deserving an Answer},
    CERN-TH.7370/94, UND-HEP-94-BIG08 (1994).
    I.I Bigi,
    {\it The Expected, The Promised and the Conceivable - on CP Violation
    in Beauty and Charm Decays.}, UND-HEP-94-BIG11 (1994).
\bibitem{d0mixing}   
    Tiehui (Ted) Liu,
    {\it The D0-Dobar Mixing Search -- Current Status and Future Prospects},
    HUTP-94/E021 (1994).
    Gustavo Burdman,
    {\it Charm Mixing and CP Violation in the Standard Model},
    FERMILAB-Conf-94/200 (1994).

\end{references}
\end{document}